# A NEW METHODOLOGY FOR THE OPTIMIZATION OF BOLT TIGHTENING SEQUENCES FOR RING TYPE JOINTS

Ibai Coria[1], Mikel Abasolo[1], Imanol Olaskoaga[2], Arkaitz Etxezarreta[2], Josu Aguirrebeitia[1]

[1] Department of Mechanical Engineering, University of the Basque Country (UPV/EHU), Bilbao, Spain.

[2] Fastening Solutions Division, Matz-Erreka, S.Coop, Bergara, Spain.

**Abstract**

Achieving uniform bolt load distribution is critical to obtain leak-free service in pressure vessel gasketed joints used in offshore pipelines. This is a difficult task due to bolt load variations during the assembly process. In this sense, the Elastic Interaction Coefficients Method has been developed in previous works to define tightening sequences that provide the target load at the end of the sequence in one or two passes. The method is very costly because a complete sequence must be simulated and the load of every bolt must be measured after each tightening operation. The present work validates this method for Ring Type Joints and further develops a numerically and experimentally validated new methodology that provides highly satisfactory results with a significantly lower cost.

## 1. Introduction

In the assembly of pressure vessel gasketed joints, uniform bolt force is critical to achieve uniform distribution of gasket stress and therefore leak-free service of the joint. Nevertheless, whenever a fastener is tightened in a gasketed joint during the assembly, the joint gets compressed and consequently the load in previously tightened fasteners is reduced. This phenomenon is known as elastic interaction or bolt cross talk (Bickford and Nassar, 1998). The magnitude of these load variations depend on a large number of parameters whose influence is hard to predict, such as geometry and material of the joint components, load magnitude, bolt spacing, assembly pattern, amongst others

(Nassar and Alkelani, 2005; Takaki and Fukuoka, 2002). Thus, obtaining a uniform load distribution at the end of the assembly is not straightforward.

Several standards contain different bolt assembly procedures which have shown good performance in terms of achieving final uniform load and therefore preventing gasket damage and reducing leakage incidents (API, 2011; ASME, 2013; Brown et al., 2010; NORSOK, 2013). In all of them, the tightening sequence is carried out in several passes gradually increasing the value of the torque applied to the bolts in each pass. The tightening sequence is generally following a star pattern or similar (circular patterns, if present, are only used for the latest passes), which ensures a better alignment of matching flanges and avoids local overloads in the gasket and rigid body motion in the joint (Bickford, 1995). However, standards point out that these sequences are indicative and generalist and they recommend that each assembler should develop his own sequences that suit their particular products and working conditions.

Significant effort in research has gone into developing faster assembly methods that provide a uniform load distribution in a single pass or in two pass sequences. This process is known as the *optimization of the tightening sequence*. The most popular method in the specialized literature is the *Elastic Interaction Coefficients Method* (EICM) (Bibel and Ezell, 1996, 1992; Fukuoka and Takaki, 2004; Van Campen, 1969), which has been validated in different gasketed joints (Bibel, 1994; Ezell, 1992), in a cylinder head of highly variable stiffness and contact geometry (Goddard and Bibel, 1994) or even in wind turbine generator flanges (Abasolo et al., 2014, 2011) among others. Analytical methods have also been developed for the optimization of the tightening sequence for joints with highly non-linear gaskets (Abid et al. 2016, 2015; Fukuoka and Takaki, 2003).

In the present work, the optimization of the tightening sequence for metallic gasketed Ring Type Joints (RTJ) (ASME, 2003) is studied. These joints are excel in offshore Oil & Gas applications because of their capacity to provide a high integrity seal at very high internal pressure (Currie, 2012). The benefit is that the gasket is confined in a groove and has two sealing surfaces. Besides it tends to be self-actuating, i.e. the sealing pressure increases with the increase of the service pressure.

As a first step, the validity of the EICM for the optimization of the tightening sequence for RTJs has been studied (sections 4 and 5 of the article). Then, as a major contribution of this work, a new methodology for the optimization of the tightening sequence for RTJs, called the *Tetraparametric Assembly Method* (TAM), has been developed, which is a further improvement of the EICM due to its significantly lower cost (section 6 of the article). Prior to that, sections 2 and 3 describe respectively the joint under study and the analysis tools used in this work (FE model and test bench).

## 2. Joint under study and operational variables

This work studies a Ring Type Joint (RTJ), composed by a pair of 24" NPS Class 150 RTJ SCHD 40 flanges (ASME, 2003) and a R76 metallic ring gasket with octagonal profile (ASME, 2012) (Fig. 1). The materials are ASTM A105 steel ($E$=201 GPa, $v$=0.3) for the flange and soft iron ($E$=198 GPa, $v$=0.285) for the gasket. The bolts size is $1^{1/4}$-8 (UN Series) Class 10.9, with tensile stress area 1 in$^2$.

In order to observe the influence of the magnitude of tightening load in the joint behavior, two values were used, 200 kN and 300 kN (approximately 40% and 70% of the yield stress of the bolts). Besides, the two assembly patterns illustrated in Fig. 2 were analyzed. The friction coefficient in the flange-gasket contact depends on many factors and it is hardly predictable; for the calculations in this work, two representative values were chosen: $\mu$=0.2 and $\mu$=0.3.

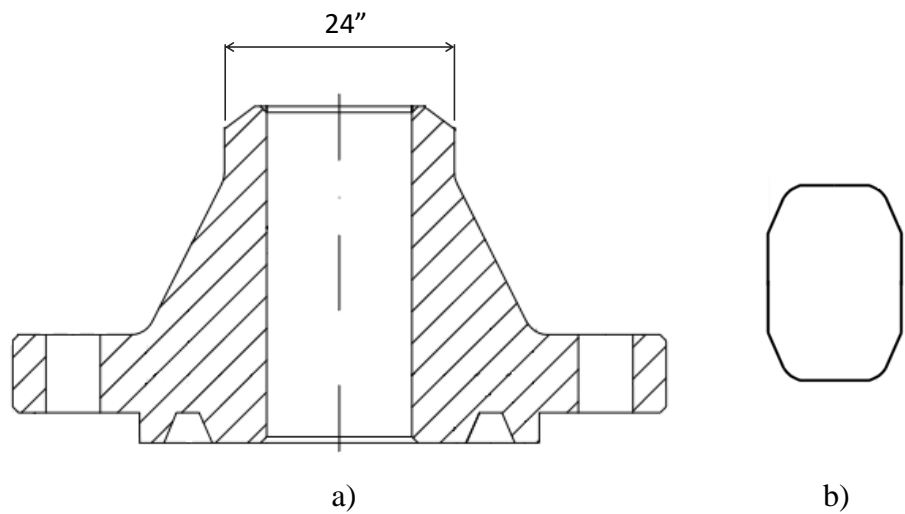

a) b)

**Fig. 1.** Studied joint: a) 24" NPS Class 150 RTJ SCHD 40 (ASME, 2003) b) R76 metallic ring gasket with octagonal profile (ASME, 2012).

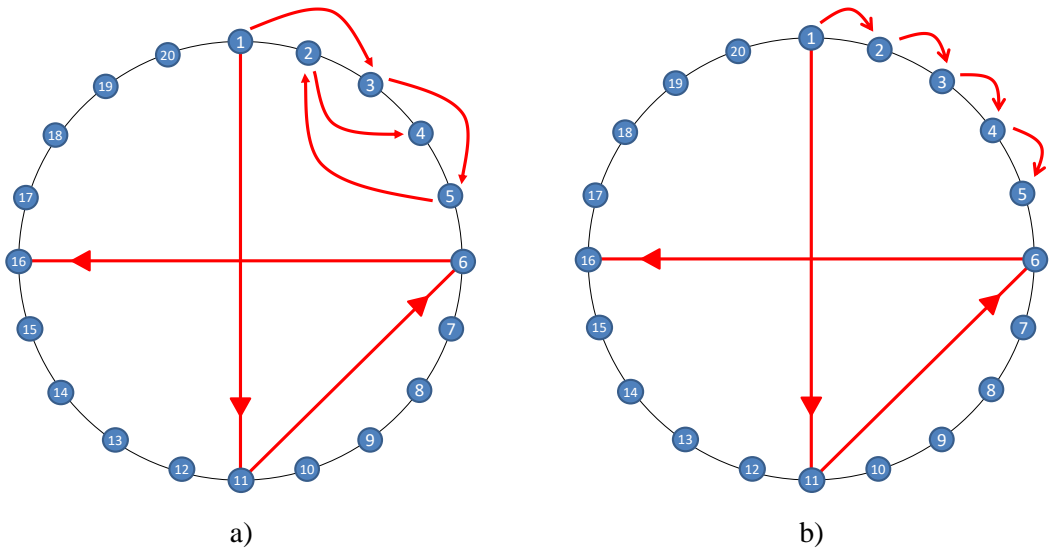

**Fig. 2.** Analyzed assembly patterns: a) Pattern 1: 1-11-6-16 → 3-13-8-18 → 5-15-10-20 → 2-12-7-17 → 4-14-9-19 b) Pattern 2: 1-11-6-16 → 2-12-7-17 → 3-13-8-18 → 4-14-9-19 → 5-15-10-20.

## 3. FE model and test bench

The optimization of the tightening sequence with the EICM and the TAM was performed with a FE model. In parallel, a test bench was built to validate the FE model and the results obtained.

*3.1. FE model*

The FE model of the joint was built in Ansys Workbench® (Fig. 3). Taking advantage of the symmetry of the system, only one flange and one half of the gasket was modeled defining boundary conditions that simulate this symmetry. Furthermore, a rough (non sliding) contact was defined between each bolt and the flange, and a frictional contact between the flange and the gasket. The tightening load of the fasteners was applied via pretension sections. All of the components (flange, gasket and fasteners), were meshed with eight-node bricks and ten-node tetrahedrons, resulting in a model with 1,237,977 DOFs.

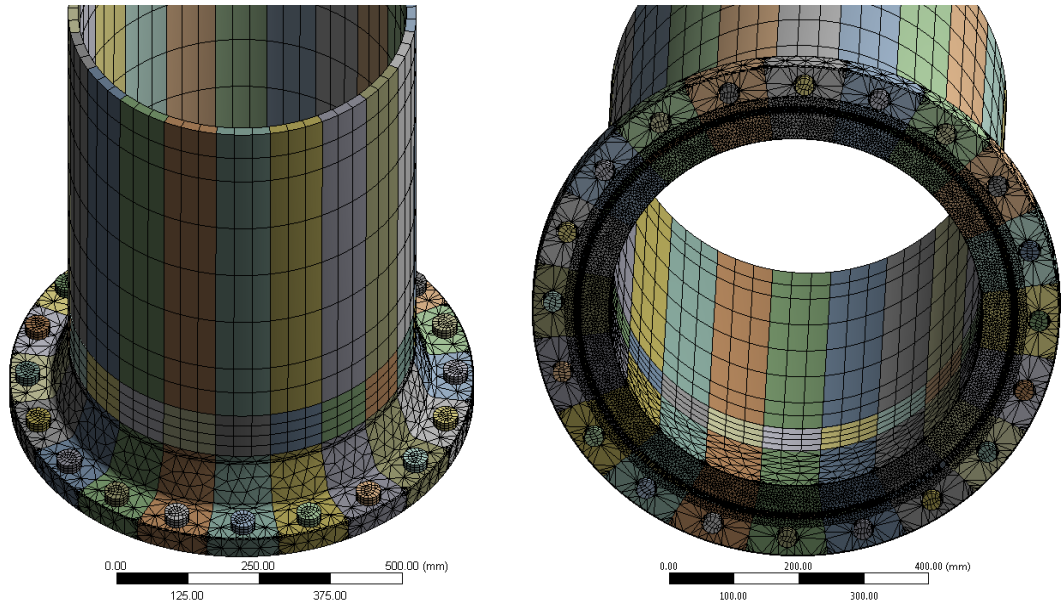

**Fig. 3.** FE model.

*3.2. Test bench*

Fig. 4 shows the test bench built for this work, composed by two pipes welded to the bolted flanges. The bolts are tightened by a hydraulic torque wrench. However, the bolt load cannot be accurately measured by measuring the torque because the torque-load relationship shows a large scatter (Bickford, 1995). To solve this important problem, the bolt load has been directly measured using i-bolt® ultrasonic measurement technology by Erreka Fastening Solutions, which provides a 3σ accuracy better than ±3% (Bickford, 1995; Erreka, 2015).

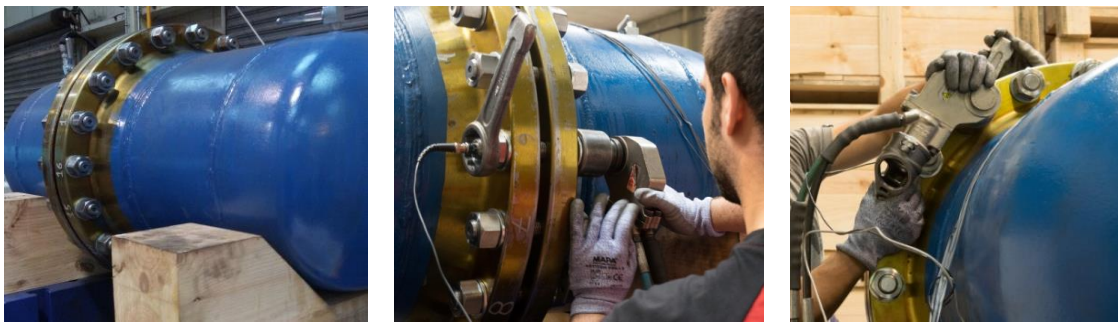

**Fig. 4.** Test bench.

*3.3. Validation of the FE model*

In order to verify that the FE model properly simulates the loss of load in the bolts during the assembly of the joint, two analyses were performed in the FE model and in the test bench. In the first analysis, pattern 1 (Fig. 2a) was followed, tightening all of the bolts to 350 kN; in the second analysis, pattern 2 (Fig. 2b) and 200 kN was applied. Fig. 5 compares the final loads obtained in both cases; FE and experimental results are similar, with average relative errors (using absolute values for the differences) of 6% and 3.5% in the first and second analyses, respectively. These results prove that the FE model can be used to measure bolt load variations with high accuracy. The small error may be caused mainly by three factors: first, the measurement error of the ultrasonic equipment; second, short-term relaxation phenomena, especially embedment of the contacting surfaces, not considered in the FE model (Bickford, 1995; Currie, 2012); and third, in the test bench a hand tightening was applied to the bolts previous to the tightening sequence, not simulated in the FE model.

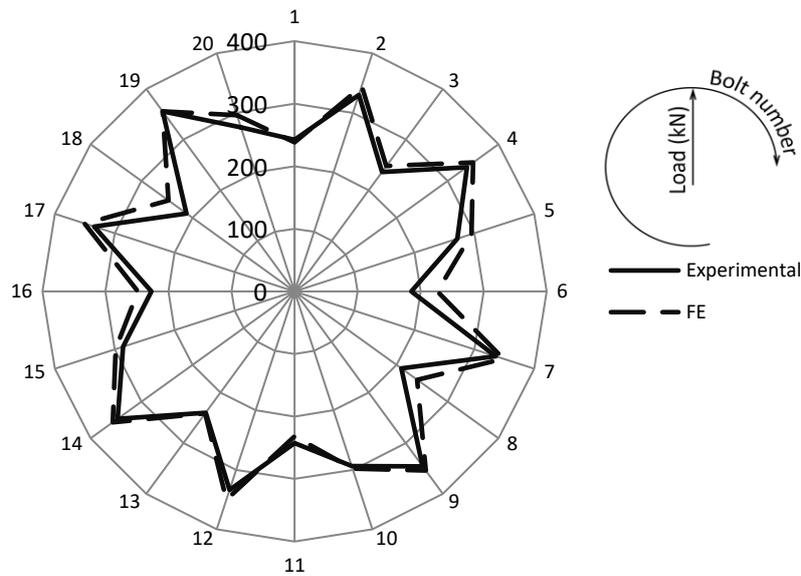

a)

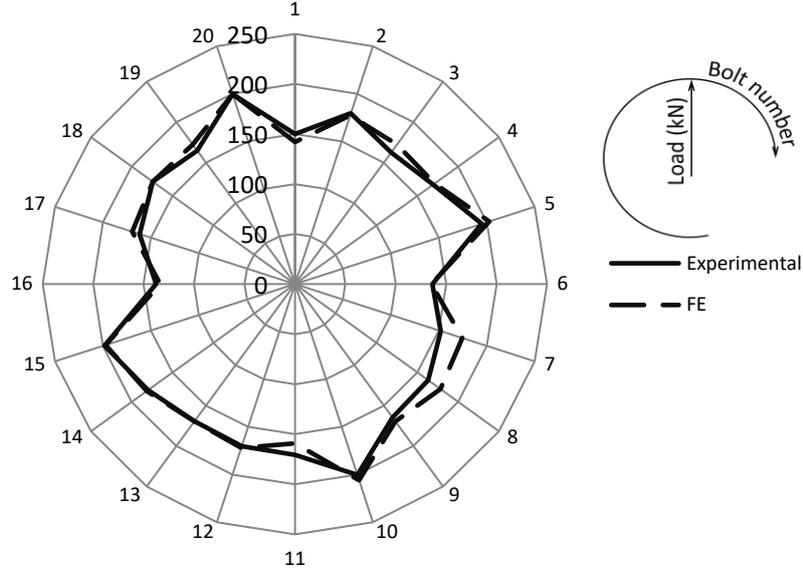

b)

**Fig. 5.** Experimental vs FE final load results: (a) tightening load 350 kN and pattern 1 (b) tightening load 200 kN and pattern 2.

## 4. Elastic Interaction Coefficients Method (EICM)

Once the joint under study and the analysis tools have been presented, the EICM will be explained. This method, which was first described by D. H. Van Campen (Van Campen, 1969) and further developed and improved by Bibel et al. (Bibel, 1994; Bibel and Ezell, 1996, 1992; Ezell, 1992; Goddard and Bibel, 1994), enables to calculate the tightening load of each bolt to achieve a target uniform load distribution at the end of the assembly in a single pass. As mentioned in the Introduction, this method serves as a basis for the TAM developed in this work.

### 4.1. Background

When defining a tightening sequence, the aim is to obtain the target load in the minimum number of passes. The EICM method is based on a matrix calculation, according to which the initial (tightening) loads and final loads (at the end of the assembly) in the bolts have the following relationship (Bibel, 1994; Bibel and Ezell, 1996, 1992; Ezell, 1992; Fukuoka and Takaki, 2004; Goddard and Bibel, 1994):

$$\{S_f\} = [A] \cdot \{S_i\} \qquad (1)$$

Where {$S_f$} and {$S_i$} are respectively the vectors with the final and initial loads in the bolts. In other words, the element $S_{ik}$ of the vector {$S_i$} is the initial load of bolt $k$, while the element $S_{fk}$ of the vector {$S_f$} is the final load of bolt $k$. Thus, in a joint composed by $n$ bolts, equation (1) has the following shape:

$$\begin{Bmatrix} S_{f1} \\ S_{f2} \\ S_{f3} \\ \vdots \\ S_{fn-1} \\ S_{fn} \end{Bmatrix} = \begin{bmatrix} 1 & A_{1,2} & A_{1,3} & \cdots & A_{1,n-1} & A_{1,n} \\ 0 & 1 & A_{2,3} & \cdots & A_{2,n-1} & A_{2,n} \\ 0 & 0 & 1 & \cdots & A_{3,n-1} & A_{3,n} \\ \vdots & \vdots & \vdots & \ddots & \vdots & \vdots \\ 0 & 0 & 0 & \vdots & 1 & A_{n-1,n} \\ 0 & 0 & 0 & \vdots & 0 & 1 \end{bmatrix} \begin{Bmatrix} S_{i1} \\ S_{i2} \\ S_{i3} \\ \vdots \\ S_{in-1} \\ S_{in} \end{Bmatrix} \quad (2)$$

It is important to note that the bolts are numbered in the order in which they are tightened; the first row corresponds to the first bolt tightened in the assembly pattern, the second row to the second bolt of the assembly pattern, and so on. If a certain row of equation (2) is studied, for instance the first row, related to the first bolt being tightened in the assembly, it is obtained that:

$$S_{f1} = S_{i1} + A_{1,2} \cdot S_{i2} + A_{1,3} \cdot S_{i3} + \ldots + A_{1,n-1} \cdot S_{in-1} + A_{1,n} \cdot S_{in} \quad (3)$$

Equation (3) indicates that the final load of bolt 1 ($S_{f1}$) is equal to its initial load ($S_{i1}$) plus $A_{1,2}$ times the initial load of bolt 2 ($S_{i2}$), plus $A_{1,3}$ times the initial load of bolt 3 ($S_{i3}$), etcetera. This means that bolt 1 has a initial load of $S_{i1}$, when bolt 2 is tightened it suffers a load variation of $A_{1,2} \cdot S_{i2}$, afterward the bolt 3 causes a load variation of $A_{1,3} \cdot S_{i3}$, and so on until the assembly is completed, giving as a result the final load $S_{f1}$. Generally, elements $A_{i,j}$ of matrix *[A]* have a negative value, since load variations are usually losses of load. It can be deducted that the only way to obtain the values of elements $A_{i,j}$ of matrix *[A]* is by simulating the whole sequence (in a FE model or in a test bench) and measuring the load variations in all of the bolts after each tightening. Once matrix *[A]* has been calculated, the initial loads {$S_i$} that must be applied to obtain a target final load distribution {$S_f$} are calculated with the following expression:

$$\{S_i\} = [A]^{-1} \cdot \{S_f\} \quad (4)$$

As explained, the EICM is based on the calculation of matrix *[A]*. Next, the algorithm for its calculation is presented with a simple illustrative example (similar examples can be found in Bibel and Ezell, 1992; Goddard and Bibel, 1994).

## 4.2. Calculation algorithm of matrix [A]

Fig. 6 shows a joint with three bolts in which the target load is 10000 N, tightening first bolt *a*, then bolt *c* and finally bolt *b*. According to the EICM, the initial load to be applied to each of the bolts is calculated with expression (4). Therefore, matrix *[A]* must be calculated.

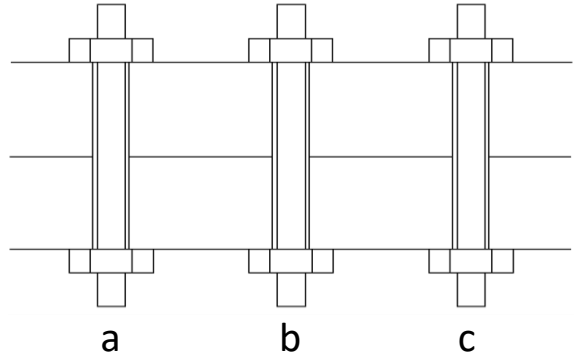

**Fig. 6.** Schematic representation of the joint of the illustrative example.

For such purpose, the EICM indicates that a tightening sequence must be simulated (in a FE model or in a test bench), using 10000 N as initial loads for all of the bolts. Assume that, having done so, when bolt *c* is tightened the load of bolt *a* decreases to 8250 N, and when bolt *b* is tightened the load of bolts *a* and *c* decrease respectively to 7500 N and 9000 N. With this data, auxiliary matrix *[$S_h$]* is built:

$$[S_h] = \begin{bmatrix} 10000 & 0 & 0 \\ 8250 & 10000 & 0 \\ 7500 & 9000 & 10000 \end{bmatrix}$$

The first column contains the loads of bolt *a* (first bolt of the assembly pattern), whereas second and third columns show the loads of bolt *c* (second bolt of the assembly pattern) and bolt *b* (third bolt of the assembly pattern), respectively. Moreover, the first row contains the loads after tightening bolt *a*, the second row the loads after tightening bolt *c*, and the third row the loads after tightening bolt *b* (i.e. at the end of the assembly pattern). Thus, matrix *[$S_h$]* stores the loads in all of the bolts during the whole tightening sequence. From this matrix, elements $A_{i,j}$ of matrix *[A]* can be calculated as (Bibel and Ezell, 1992; Goddard and Bibel, 1994):

$$\begin{aligned} A_{i,j} &= 1 & i = j \\ A_{i,j} &= 0 & i > j \\ A_{i,j} &= \frac{[S_h]_{j,i} - [S_h]_{j-1,i}}{[S_h]_{j,j}} & i < j \end{aligned} \qquad (5)$$

In the example, operating:

$$[A] = \begin{bmatrix} 1 & -0.175 & -0.075 \\ 0 & 1 & -0.1 \\ 0 & 0 & 1 \end{bmatrix}$$

As explained in the previous section, the first row indicates that the final load of bolt *a* is equal to his tightening load, minus 0.175 times the tightening load of bolt *c*, minus 0.075 times the tightening load of bolt *b*. By using equation (4), the initial loads to obtain the target final load distribution can be easily calculated.

*4.3. Discussion on the method*

In conclusion, in order to calculate matrix *[A]* the whole sequence must be simulated and the load of every bolt after each tightening operation must be measured, in a FE model or a test bench; obviously, this is a very laborious process. Moreover, equation (2) assumes that the relationship between the initial and final loads in the bolts is completely linear; however, the response of the joint during the assembly can be non-linear due to non-linear gasket materials, changes in the contact status between components or large deflections. In that case, matrix *[A]* must be iteratively calculated in order to suit to the working loads, which substantially increases the work to be performed (Bibel, 1994; Bibel and Ezell, 1994, 1996; Ezell, 1992).

Next, as commented in the Introduction, the validity of the EICM for the joint under study is investigated, also assessing if matrix *[A]* has to be calculated iteratively. A deep analysis of the results has led to the development of a new optimization methodology (TAM) that will be explained afterwards.

**5. Validation of the Elastic Interaction Coefficients Method for the joint under study**

From the FE results obtained in section 3.3, matrix *[A]* was calculated, and the initial loads to be applied to the bolts were determined according to the EICM for two cases: the first one, to obtain a final uniform load of 350 kN using pattern 1; the second one, with a target load of 200 kN using pattern 2. Fig. 7 shows the initial load values according to the EICM and the final loads obtained from the FE simulations: the resulting final load distribution was completely uniform, with average loads of 347 kN and 200 kN, and standard deviations of 4.7 kN and 0.4 kN, respectively. This proved that the EICM provides very satisfactory results for the joint studied, with no need to calculate iteratively matrix *[A]* in spite of the non-linear behavior of the joint due to the contact between the flange and the gasket (Currie, 2012).

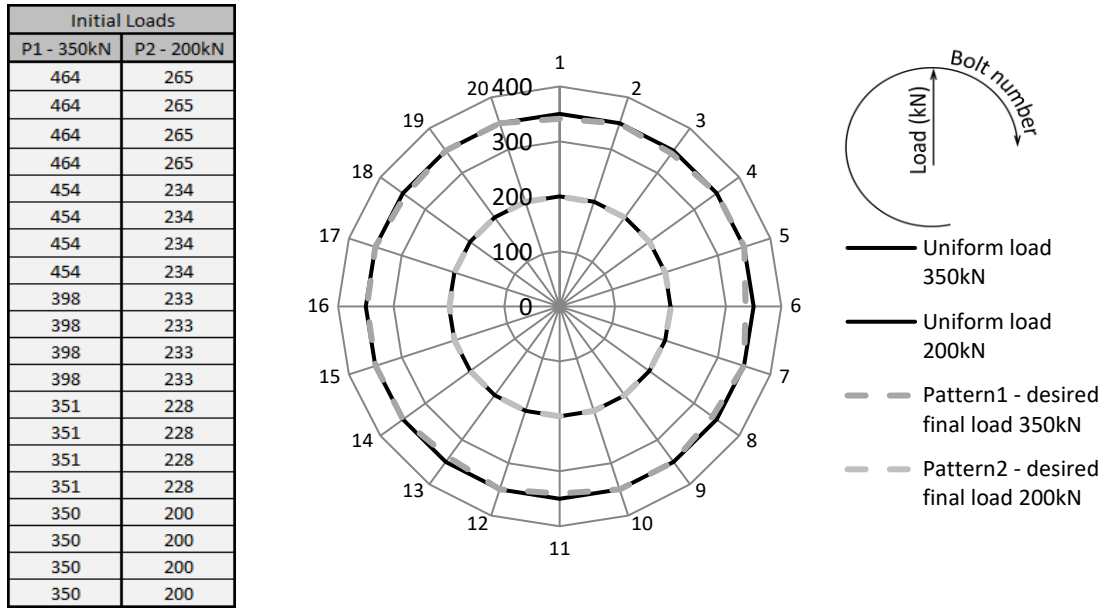

**Fig. 7.** Results for the validation of the EICM for RTJ flanges.

## 6. New methodology: Tetraparametric Assembly Method (TAM)

*6.1. Study of the elements of matrix [A]*

Once the EICM was validated, all of the cases mentioned in section 2 were analyzed in order to estimate the influence of the friction coefficient, the load magnitude and the pattern in matrix *[A]*. The conclusions were used further on to lay the ground of the TAM.

Fig. 8 shows the shape of the calculated *[A]* matrices. As it has been explained in section 4, coefficients $\alpha$, $\beta$, $\gamma$ and $\delta$ quantify the interaction between bolts during the assembly process. The position of those coefficients in rows and columns in matrix *[A]* is different for pattern 1 and pattern 2, since the bolts are tightened following a different order; nevertheless, as it will be explained in the next section, these positions are easily predictable for different patterns. Table 1 shows the values for the coefficients for each of the performed analyses, from which the following conclusions can be drawn: the values are almost identical for both load magnitudes (200 kN and 350 kN), and very similar for both patterns (pattern 1 and pattern 2) and friction coefficients (0.2 and 0.3). Therefore, it can be stated that a certain bolted joint has a unique matrix *[A]*, composed of coefficients $\alpha$, $\beta$, $\gamma$ and $\delta$ that are practically identical regardless of the friction coefficient and load magnitude (δ differs considerably from one case to another, but its contribution is almost negligible since its value is much lower than the rest of the coefficients); for different patterns, the coefficients occupy a different but predictable position in matrix *[A]*. This involves that matrix *[A]* for any joint can be obtained for a single simulation using any operational conditions (friction coefficient, load magnitude and pattern).

Based on this conclusion, a new methodology was developed that enabled to obtain matrix *[A]* with no need to simulate a complete sequence and to measure the bolt loads after every tightening step, as required by the EICM (see section 4). Thus, the new methodology, called the *Tetraparametric Assembly Method* (TAM), significantly minimizes the cost of the optimization procedure.

**Fig. 8.** Matrix *[A]* of joint under study: (a) pattern 1 (b) pattern 2.

**Table 1.** Values of coefficients of matrix *[A]* according to the EICM.

| Load | μ | Pattern 1 | | | | Pattern 2 | | | |
|---|---|---|---|---|---|---|---|---|---|
| | | α | β | γ | δ | α | β | γ | δ |
| 200 | 0.2 | -0.147 | -0.147 | -0.018 | 0.002 | -0.152 | -0.151 | -0.019 | 0.002 |
| | 0.3 | -0.139 | -0.138 | -0.019 | -0.002 | -0.144 | -0.142 | -0.020 | -0.001 |
| 350 | 0.2 | -0.148 | -0.148 | -0.021 | 0.000 | -0.155 | -0.153 | -0.019 | 0.001 |
| | 0.3 | -0.140 | -0.139 | -0.022 | -0.004 | -0.147 | -0.144 | -0.021 | -0.003 |

*6.2. Development of the Tetraparametric Assembly Method (TAM)*

Most elements in matrices *[A]* in Fig. 8 are null. This means that when a bolt is tightened during the assembly process, only a few bolts experience loss of load (obviously, the bolts closest to the one being tightened). For example, bolt 1 (first row of the matrix) only loses load when bolts 2, 3, 19 or 20 are tightened, i.e. the bolts located in one or two positions from bolt 1. In other words, the tightening of one bolt only affects to the bolts located to one or two positions of distance, while the remaining bolts have no load variation. Therefore, coefficients *α*, *β*, *γ* and *δ* of matrix *[A]* in Fig. 8 correspond to the four possible load cases which can arise during the tightening sequence. It must be remarked that in other more flexible RTJ geometries, more bolts could be affected; the validation of the new methodology for such cases remains as a further research work. Table 2 explains schematically each of these load cases and the formulas to work out the coefficients, which have been directly deducted from equation (5). It must be emphasized that a RTJ joint much more flexible than the one under study would need more coefficients due to a larger elastic interaction. Nevertheless, the procedure presented can easily be generalized for that case.

Having understood the meaning of the coefficients, matrices *[A]* in Fig. 8 can be analyzed. For example, row 1 of matrix *[A]* in Fig. 8a computes the loss of load of bolt 1 during pattern 1 (Fig. 2a). In this sense, according to Table 2: *γ* occupies the fifth column because when bolt 3 is tightened, bolt 2 has not been tightened yet; *α* occupies the twelfth column because when bolt 20 is tightened, bolt 19 has not been tightened yet; *β* occupies the thirteenth column because when bolt 2 is tightened, bolt 3 has been previously tightened; *δ* occupies the twentieth column because when bolt 19 is tightened, bolt 20 has been previously tightened. In Fig. 8b, the coefficients occupy different positions because the tightening pattern is different (pattern 2, Fig. 2b), but the same procedure is applied, so these positions are easily predictable. This proves that for different patterns, matrix *[A]* is the same but with the coefficients in different rows and columns, as it was stated in the previous section.

**Table 2.** Load cases for coefficients of matrix *[A]*. Formulas derived from equation (5).

| | | |
|---|---|---|
| b<br>b+1<br>b+2 | Bolt *b* is tightened to load $F_b$.<br>Bolt *b+1* was previously tightened to load $F_{b+1}$ and, after tightening bolt *b*, its load becomes $F_{b+1}'$.<br>Bolt *b+2* is not previously tightened. | $\alpha = \dfrac{F_{b+1}' - F_{b+1}}{F_b}$<br><br>$\alpha$ estimates the loss of load of bolt *b+1* when bolt *b* is tightened, with bolt *b+2* not previously tightened. |
| b<br>b+1<br>b+2 | Bolt *b* is tightened to load $F_b$.<br>Bolt *b+1* was previously tightened to load $F_{b+1}$ and, after tightening bolt *b*, its load becomes $F_{b+1}'$.<br>Bolt *b+2* was previously tightened to load $F_{b+2}$ and, after tightening bolt *b*, its load becomes $F_{b+2}'$. | $\beta = \dfrac{F_{b+1}' - F_{b+1}}{F_b}$<br><br>$\beta$ estimates the loss of load of bolt *b+1* when bolt *b* is tightened, with bolt *b+2* previously tightened.<br><br>$\delta = \dfrac{F_{b+2}' - F_{b+2}}{F_b}$<br><br>$\delta$ estimates the loss of load of bolt *b+2* when bolt *b* is tightened, with bolt *b+1* previously tightened. |
| b<br>b+1<br>b+2 | Bolt *b* is tightened to load $F_b$.<br>Bolt *b+1* is not previously tightened.<br>Bolt *b+2* was previously tightened to load $F_{b+2}$ and, after tightening bolt *b*, its load becomes $F_{b+2}'$. | $\gamma = \dfrac{F_{b+2}' - F_{b+2}}{F_b}$<br><br>$\gamma$ estimates the loss of load of bolt *b+2* when bolt *b* is tightened, with bolt *b+1* not previously tightened. |

The concepts outlined in Table 2 are not only useful to understand the composition of matrix *[A]*, they have also been used as the basis for the development of the TAM that allows to calculate matrix *[A]* using only two simple load steps. In the first load step several bolts are tightened to a certain load; as previously mentioned, the magnitude of the load is not relevant, although it is recommended to use a level of load similar to the target final load; similarly, in case of using a FE model the friction coefficient should be as realistic as possible, although this parameter does not have a major influence either.

In the second load step, a few other bolts are tightened, in such a way that all of the situations outlined in Table 2 take place; thus, measuring the load variation in the bolts tightened in the first load step, coefficients $\alpha$, $\beta$, $\gamma$ and $\delta$ of matrix *[A]* are calculated using equations in Table 2.

Fig. 9 shows schematically the two load steps for the particular 24" NPS joint studied in this work: the bolts tightened in the first load step are marked with a circle and the bolts of the second load step with a triangle. According to the Table 2, $\alpha$ is calculated from the loss of load of bolt 8 when bolt 9 is tightened, or from the loss of load of bolt 15 when bolt 16 is tightened; $\beta$ is calculated from the loss of load of bolt 4 when bolt 3 is tightened, or from the loss of load of bolt 17 when bolt 16 is tightened; $\gamma$ is calculated from the loss of load of bolt 1 when bolt 3 is tightened, or from the loss of load of bolt 11 when bolt 9 is tightened; finally, $\delta$ is calculated from the loss of load of bolt 5 when bolt 3 is tightened, or from the loss of load of bolt 18 when bolt 16 is tightened.

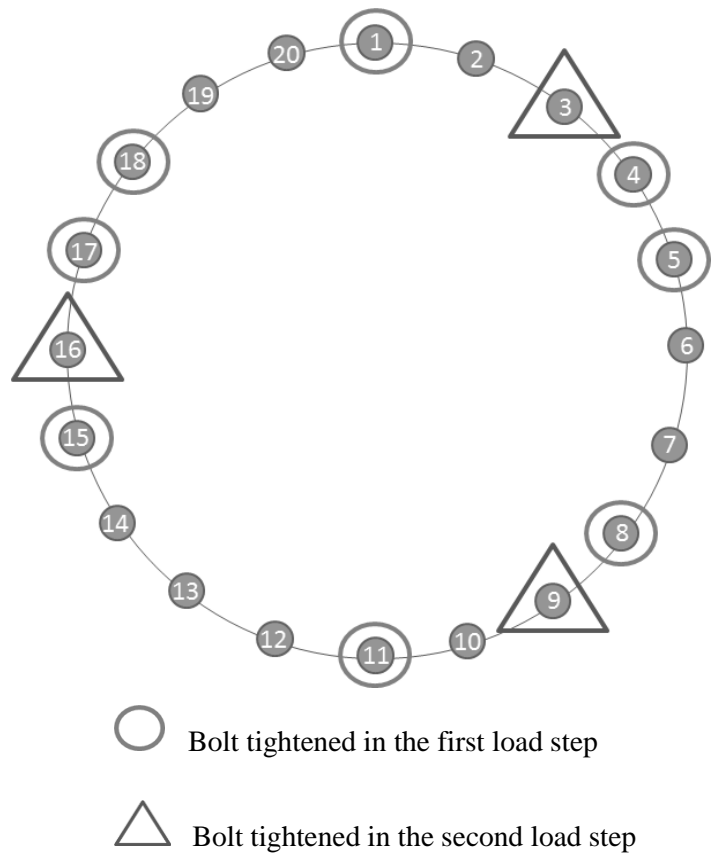

○ Bolt tightened in the first load step

△ Bolt tightened in the second load step

**Fig. 9.** Two load steps to obtain matrix *[A]* using the TAM.

Obviously, the TAM has a much lower cost than the EICM. If a FE model is to be used, the TAM needs two load steps, while the EICM needs twenty load steps (there are twenty bolts in the joint) because the whole sequence must be simulated; besides, simulating the sequence may present convergence problems due to the instability of the joint in the first load steps when only a few bolts are loaded, which further increases its cost. If a test bench is used, as it has just been explained, the TAM needs eleven tightening operations and nineteen load measurements, whereas the EICM simulates the whole sequence and measures the bolt loads of previously tightened bolts after each tightening, resulting in twenty tightening operations and 210 load measurements. Next, the results provided by the TAM are presented and compared with the results of the EICM.

*6.3. Results and discussion*

Table 3 presents the values of the coefficients calculated using the TAM. The results are very similar to the ones presented in Table 1 (obtained from the simulations of the complete sequence as required by the EICM), proving thus the validity of the TAM.

**Table 3.** Values of coefficients of matrix *[A]* according to the TAM.

| $\mu$ | $\alpha$ | $\beta$ | $\gamma$ | $\delta$ |
|---|---|---|---|---|
| 0.2 | -0.147 | -0.147 | -0.018 | 0.002 |
| 0.3 | -0.139 | -0.138 | -0.019 | -0.002 |

A further validation consists on computing the initial loads obtained with the TAM (coefficients in Table 3) and with the EICM (coefficients in Table 1). These initial loads, calculated using equation (4), are presented in Fig. 10 as a ratio between initial load and final load. It may be observed that the results of the EICM and of the TAM are highly coincident (always with a relative error less than 2% for every value), as might have been expected being coefficients $\alpha$, $\beta$, $\gamma$ and $\delta$ of the matrix *[A]* so similar.

It can be observed in Fig. 10 that there is a slight difference in the results for different friction coefficients, because the values of coefficients in matrix $\alpha$, $\beta$, $\gamma$ and $\delta$ are different (discrepancy around 5%). However, the difference is small, so that it can be stated that the friction coefficient does not have a significant effect; nevertheless, in order to obtain accurate values for $\alpha$, $\beta$, $\gamma$ and $\delta$, it is recommended that the two step analysis in Figure 9 should be performed experimentally instead of with FEA.

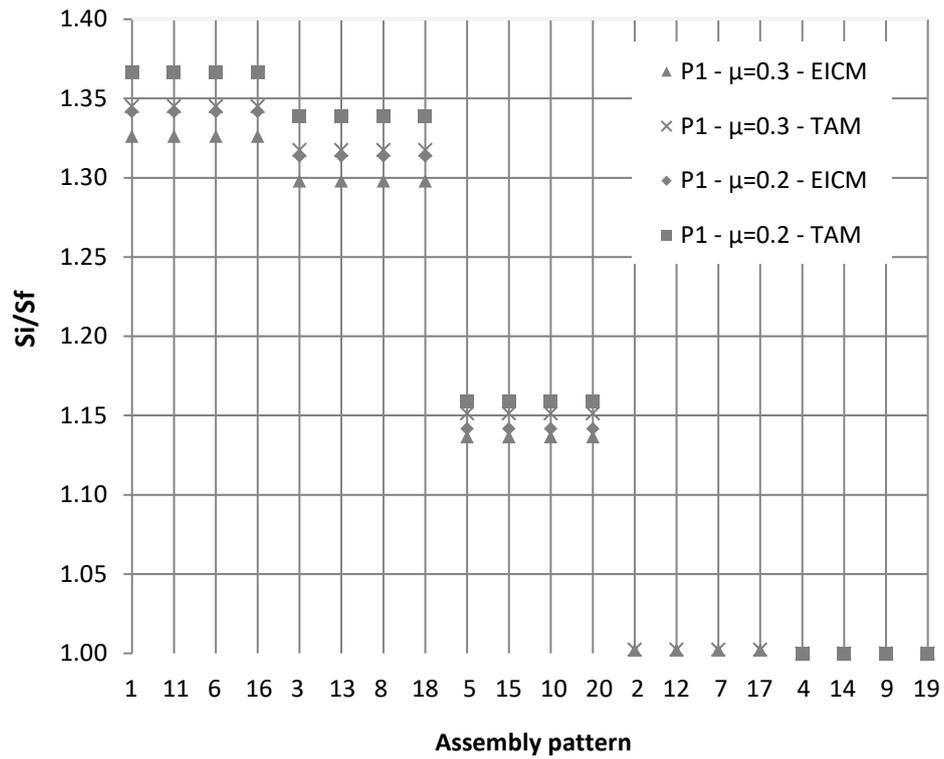

a)

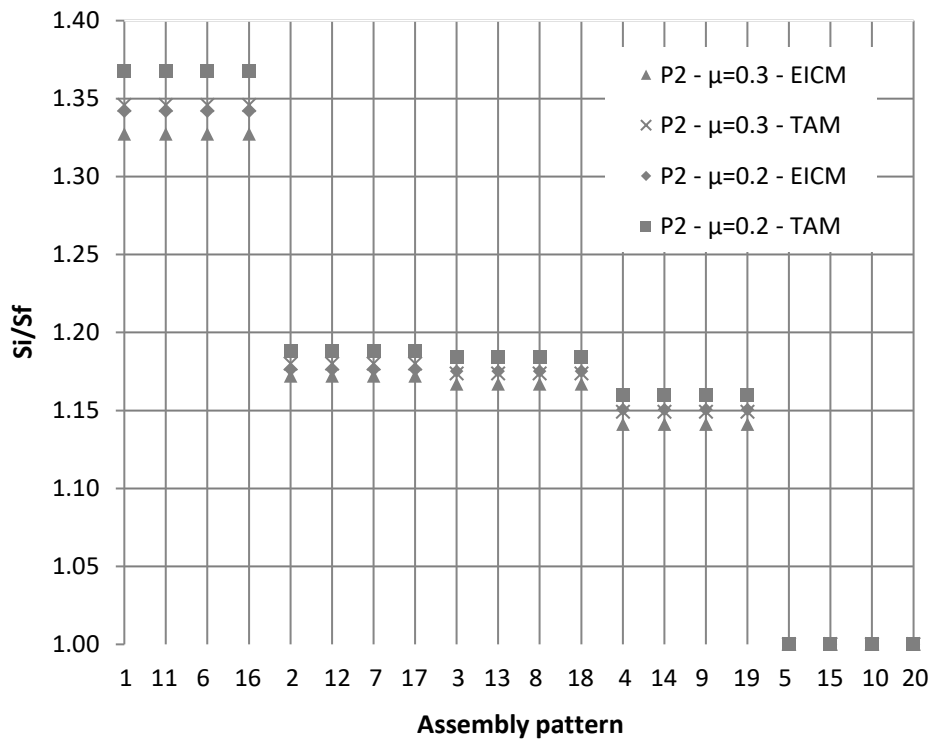

b)

**Fig. 10.** Ratio between initial loads and final loads to obtain a uniform final load distribution:
(a) pattern 1 (b) pattern 2.

The initial loads obtained by the TAM were applied to the bolts in the test bench. In the first test, pattern 1 with initial loads in Fig. 11 (obtained from Fig. 10a) was applied, to obtain a target final load of 200 kN. The average load achieved was 207 kN with a standard deviation of 6.6 kN (part of this error was because the tightening loads introduced by the torque wrench were slightly different from those indicated in Fig. 11).

Another similar test was carried out using a new tightening pattern, in which the first four bolts were tightened following a star pattern and then the remaining bolts in a circular pattern (star-circular pattern); from the point of view of the assembly operation, this pattern is much simpler and faster than those proposed in Fig. 2. The initial loads calculated with the TAM to obtain a uniform load distribution of 200 kN and the final loads obtained in the test bench are shown in the Fig. 12; once again, the results are highly satisfactory, with an average load of 198 kN and a standard deviation of 5.1 kN.

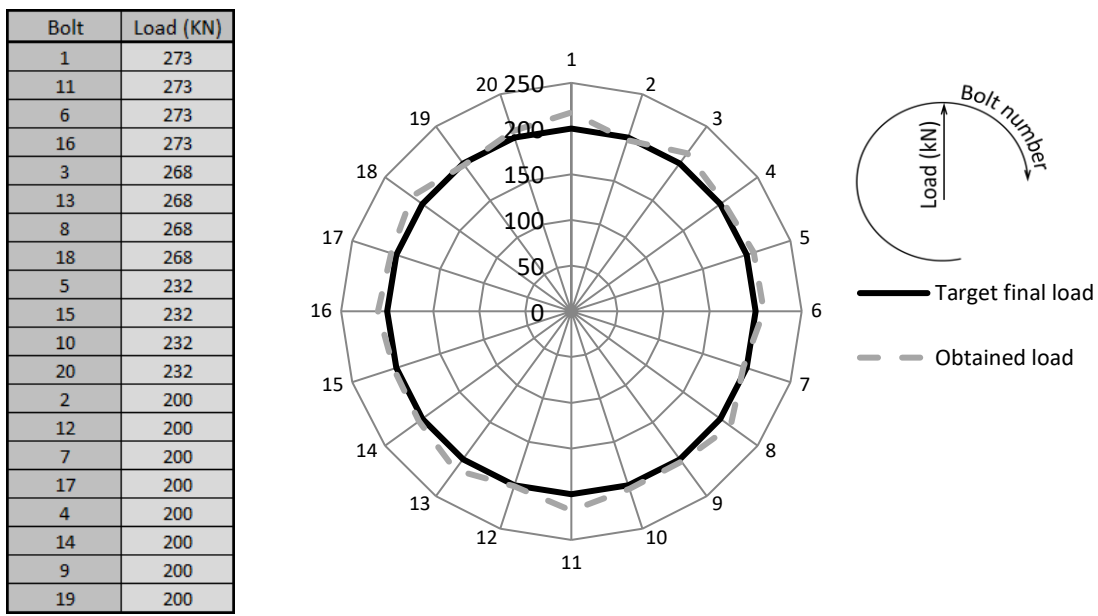

**Fig. 11.** Experimental results of the optimization process using the TAM (target load 200 kN and pattern 1).

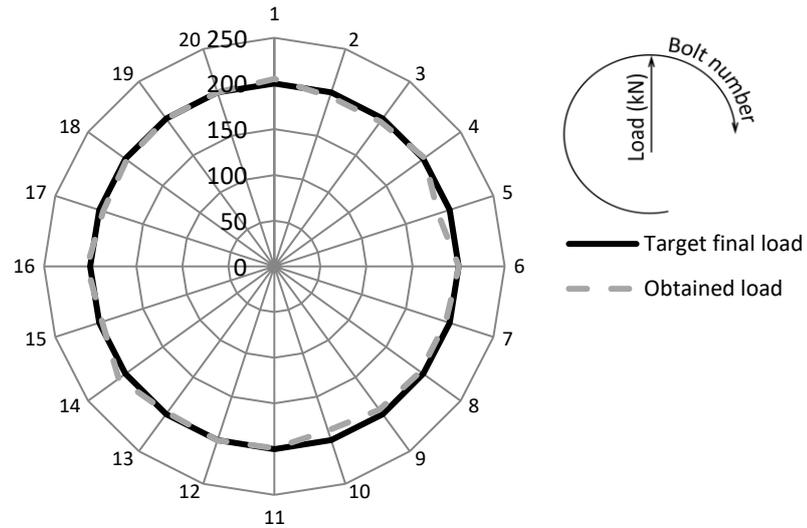

| Bolt | Load (KN) |
|------|-----------|
| 1 | 273 |
| 11 | 274 |
| 6 | 274 |
| 16 | 274 |
| 2 | 238 |
| 3 | 237 |
| 4 | 232 |
| 5 | 200 |
| 7 | 238 |
| 8 | 237 |
| 9 | 232 |
| 10 | 200 |
| 12 | 238 |
| 13 | 237 |
| 14 | 232 |
| 15 | 200 |
| 17 | 238 |
| 18 | 237 |
| 19 | 232 |
| 20 | 200 |

**Fig. 12.** Experimental results of the optimization process using the TAM (target load 200 kN and star-circular pattern).

The results show the great accuracy of the TAM for the assembly of the Ring Type Joint under study in this work (see section 2). As it was mentioned previously, further tests are necessary to prove its validity in other RTJ geometries (nominal pipe size, class and schedule).

Finally, it must be noted that the possible yield of the bolts due to an excessive initial load has not been considered in this work. In that case the tightening sequence should consist of two passes, which means that two different matrices would be necessary, one for each pass (Bibel and Ezell, 1996; Ezell, 1992; Fukuoka and Takaki, 2004). The matrix for the first pass can be obtained using the TAM presented in this manuscript; further research by the authors is focusing on adapting the TAM for the calculation of the matrix for the second pass.

Similarly, large bolt loads can cause local yielding of the gasket. Nevertheless, it has been verified that the effect of this phenomenon on the elastic interaction is not relevant. Thus, for the sake of simplicity and cost-efficiency, the coefficients for the TAM have been obtained assuming elastic material in the two step analysis in Fig. 9; the experimental results in Fig. 11 and 12 confirm that this simplification does not compromise the accuracy of the methodology. However, as it was previously mentioned for the friction coefficient, the analysis in Fig. 9 should be carried out experimentally instead of via FEA if more accurate results are wanted.

## 7. Conclusions and further research

Different standards recommend several tightening sequences for pressure vessel bolted joints, but they have a generalist nature and require several passes. In this sense, the Elastic Interaction Coefficients Method has widely been used to calculate the tightening loads for each bolt in such a way that the target uniform final load can be achieved in only one or two passes. According to it, these tightening and final loads are linearly related by means of a matrix. However, this method is costly because the calculation of the matrix requires simulating a complete tightening sequence in a FE model or in a test bench and measuring the load of every previously loaded bolt after each tightening operation. Moreover, for joints with non-linear behavior, the matrix needs to be worked out iteratively increasing thus the cost of the method.

In this work the Elastic Interaction Coefficients Method was validated for metallic gasketed Ring Type Joint (RTJ) flanges. Besides, the analysis of the results has enabled to understand the influence of the different operational conditions (load magnitude, friction coefficient and tightening pattern) in the matrix, deducting that its composition for each RTJ joint is almost invariable regardless of the operational conditions. From this conclusion a new methodology called the Tetraparametric Assembly Method (TAM) was developed for this specific type of joint, which calculates the tightening loads by using only two load steps and a few number of load measurements, thus presenting a much lower cost than the Elastic Interaction Coefficients Method. Introducing the tightening loads obtained with the TAM in a test bench, satisfactory results were achieved for the joint under study.

Regarding future work, the authors are working on three research lines. The first one is the generalization of the TAM for two pass sequences. The second one is to prove the validity of the TAM for different RTJ types (NPS, Class, SCHD) in order to fix its application range; in this sense, the authors have already tested the TAM on different NPS and SCHD values with promising results. And finally, taking advantage of the unchanging nature of the matrix for each RTJ, along with the efficiency of the TAM, the authors believe that a matrix library can be generated for assembly process standardization purposes.

Another future research consists on the adaption of the methodology for other types of pressure vessel bolted joints, for example spiral-wound gasketed joints. In these joints, the gasket is much more flexible than the flange, and thus it seems that the methodology will have to be adapted to represent the behavior of the joint during the tightening sequence.

**Acknowledgements**

The authors wish to acknowledge the financial support of the Spanish Ministry of Economy and Competitiveness (MINECO) through Grant Number DPI2013- 41091-R and the University of the Basque Country (UPV/EHU), Grant Number UFI11/29.

The authors also wish to thank Ulma Piping for making the experimental setup and facilities available.